\documentclass{article}

\usepackage{PRIMEarxiv}

\usepackage[utf8]{inputenc} 
\usepackage[T1]{fontenc}    
\usepackage{hyperref}       
\usepackage{url}            
\usepackage{booktabs}       
\usepackage{amsfonts}       
\usepackage{nicefrac}       
\usepackage{microtype}      
\usepackage{lipsum}
\usepackage{fancyhdr}       
\usepackage{graphicx}       
\graphicspath{{media/}}     

\usepackage{amsmath}
\usepackage{enumerate}

\title{Robust Phase Retrieval with Complexity-Guidance for Coherent X-ray Imaging}

\author{ 
Mansi Butola\\
  Department of Physics\\
  Indian Institute of Technology Delhi\\
  New Delhi 110016 India\\
  \texttt{mansibutola83@gmail.com} \\
   \And
  Sunaina Rajora\\
  Department of Physics\\
  Indian Institute of Technology Delhi\\
  New Delhi 110016 India\\
  \texttt{sunainarajora1511@gmail.com} \\
  \And
  Kedar Khare\\
  Optics and Photonics Centre \& \\
  Department of Physics\\
  Indian Institute of Technology Delhi\\
  New Delhi 110016 India\\
  \texttt{kedark@physics.iitd.ac.in} \\
}

\begin{document}
\maketitle

\begin{abstract}
Reconstruction of a stable and reliable solution from noisy incomplete Fourier intensity data recorded in a coherent X-ray imaging (CXI) experiment is a challenging problem for iterative phase retrieval methods. The Relaxed Averaged Alternating Reflections (RAAR) algorithm that is concluded with a number of Error Reduction (ER) iterations is a popular choice in CXI literature. The RAAR-ER algorithm is usually employed for several hundreds of times starting with independent random guesses to obtain trial solutions, that are then averaged. The resolution of the averaged reconstruction is then assessed by plotting the phase retrieval transfer function (PRTF). In this paper, we examine the phase retrieval solution obtained using the RAAR-ER methodology from perspective of the complexity parameter that was introduced by us in recent works. The complexity parameter is a measure of fluctuations in the desired phase retrieval solution and may be computed directly from the Fourier intensity data. We observe that a single run of the RAAR-ER algorithm produces a solution with higher complexity compared to what is expected based on the complexity parameter as manifested by high frequency grainy artifacts in the solution. These spurious features whose size is smaller than the predicted resolution (as per the PRTF) do not seem to go away completely even after a number of trial solutions are averaged. We then describe a CG-RAAR (Complexity Guided RAAR) phase retrieval method that can effectively address this inconsistency problem and provides artifact-free solutions by controlling the solution complexity in the RAAR iterations. The CG-RAAR methodology is first illustrated with simulated unblocked noisy Fourier intensity data and later applied to centrally blocked noisy cyanobacterium data which is available from the CXIDB database. Our simulation and experimental results using CG-RAAR suggest two important improvements over the popular RAAR-ER algorithm. First, the CG-RAAR solutions after the averaging procedure is more reliable in the sense that it contains smallest features consistent with the resolution estimated by the PRTF curve. Secondly, since the single run of the CG-RAAR solution does not have grainy artifacts, the number of trial solutions needed for the averaging process is reduced. The data domain error and PRTF based resolution estimated for the CG-RAAR solution are comparable (although slightly worse) to the traditional RAAR-ER method. This is expected due to regularizing nature of the CG-RAAR algorithm that prevents over-fitting of the solution with noise.
\end{abstract}

\keywords{Coherent X-ray imaging, Phase retrieval, Noisy diffraction data, sparsity, single particle imaging.}

\section{Introduction}
Imaging of micro and nano-sized objects remains one of the prime interests in structural biology, material science, chemistry, medical science etc. With the rapid advancements in coherent diffraction imaging hardware, particularly in X-ray free electron lasers (XFELs) \cite{bostedt2016linac}, it is possible to perform single particle imaging (SPI)
~\cite{chapman2006femtosecond, chapman2010coherent}. In order to determine the structure of objects of interest, the acquisition of Fourier diffraction data is a challenging task that is being carried out by a number of dedicated instruments \cite{struder2010large, emma2010first, liang2015coherent}. On the other hand reconstruction of robust and reliable solution from an incomplete noisy diffraction data acquired by a typical CXI experiment remains an equally important and fundamental task which is addressed by phase retrieval algorithms \cite{fienup1978reconstruction, fienup1982phase, elser2003phase, marchesini2003x, luke2004relaxed, guo2015iterative}. Since  uniqueness of the solution from a single coherent X-ray diffraction data frame cannot be guaranteed by phase retrieval algorithms, the standard procedure for the solution recovery \cite{shapiro2005biological} requires running the algorithm over hundreds of times starting with independent random guesses to get multiple trial solutions that are finally averaged together. The resolution of this averaged solution is estimated using the phase retrieval transfer function (PRTF), which is the ratio between the Fourier magnitude of the average solution and the measured Fourier magnitude as a function of the spatial frequencies. 

In recent works \cite{butola2019phase, butola2021complexity}, we have proposed a novel approach that we call ``complexity guided phase retrieval'' (CGPR) that is meant to address the typical stagnation problems with phase retrieval algorithms. This methodology uses a complexity parameter which is computed directly from the Fourier intensity data and provides a measure of fluctuations in the desired phase retrieval solution. As observed in \cite{butola2019phase, butola2021complexity}, the complexity parameter can be used to guide the application of constraints in the object domain and provides artifact free phase retrieval solution which degrades benignly with increasing noise. In \cite{butola2019phase, butola2021complexity}, the CGPR methodology was mainly developed with simulated noisy data in combination with the Fienup's hybrid input-output (HIO) algorithm. The complexity guidance idea is put to test for the first time in this work for experimental data obtained from the CXIDB database. 

Our first aim is to examine the nature of the phase retrieval solution obtained using the popular RAAR-ER methodology (both single run and averaged solution) from the perspective of the complexity parameter. A single run of RAAR-ER uses a number of relaxed averaged alternating reflection (RAAR) iterations \cite{luke2004relaxed}, that are concluded with a smaller number of error-reduction (ER) iterations for the purpose of stabilizing the solution. We observe that both single run and averaged solutions from RAAR-ER methodology consist of grainy features that are smaller in size compared to the resolution as predicted by the PRTF. Such high frequency features are difficult to interpret \cite{van2015imaging} and therefore likely to be considered as `spurious'. 

In order to address this inconsistency, we add the complexity guidance component to the RAAR algorithm to present what we call as complexity guided RAAR (CG-RAAR) algorithm. CG-RAAR is first tested with simulated noisy data (with two noise levels) that does not have missing pixels for providing a clear understanding of the proposed methodology. The CG-RAAR algorithm is then applied to the real cyanobacterium diffraction data from the CXIDB database \cite{van2015imaging} to further validate our results. The experimental data is noisy and also has missing pixels. We report two main observations. First of all a single run of CG-RAAR provides a solution without the high frequency grainy artifacts that are typically present in the RAAR-ER solution. A further consequence of this is that the number of independent runs required for averaging is reduced significantly. Since CG-RAAR essentially provides a regularized solution as guided by the complexity parameter. The CG-RAAR solution does not contain spurious grainy features.  

In view of the rapid advancements in coherent X-ray imaging (CXI) experiments, the phase retrieval algorithms are critical for obtaining a reliable solution. Recent development of open-source software called Hawk \cite{maia2010hawk}  provides the platform for the entire process of solution reconstruction from the oversampled diffraction data and creating high performance algorithms. The Hawk software implements hybrid input output (HIO) \cite{fienup1982phase}, difference map (DM) \cite{elser2003phase}, and RAAR as phasing algorithms. From our prior work in \cite{butola2021complexity} and from our experimental and simulation results demonstrated here it is evident that complexity guidance idea when incorporated to traditional phase retrieval algorithms such as HIO and RAAR offers a better noise-robust estimate of the object. We believe that complexity-guidance as an idea may potentially be integrated into existing software tools and can improve the performance of existing phasing algorithms in coherent X-ray imaging.  

\section{Methods}
In a typical coherent X-ray imaging experiment, where X-ray beam illuminates the object or sample, the Fourier intensity $I(\boldsymbol{u})$ measured at the detector can be mathematically expressed as:
\begin{align}
    I(\boldsymbol{u}) = |\hat{\rho}(\boldsymbol{u})|^2 = |\mathcal{F} \{\rho(\boldsymbol{r}) \}|^2,
\end{align}
where, $\boldsymbol{r} = (x, y)$ and $\boldsymbol{u} = (f_x, f_y)$ are the spatial and Fourier space co-ordinates. $\rho(\boldsymbol{r})$ is the electron density of the object and $\hat{\rho}(\boldsymbol{u})$ is the Fourier transform of $\rho(\boldsymbol{r})$. Through out the paper the hat notation will be the representative of Fourier domain. The aim of iterative phase retrieval algorithms is to find $\rho(\boldsymbol{r})$ from the measured intensity $I(\boldsymbol{u})$. Starting with a random guess $\rho_0(\boldsymbol{r})$, the Fourier magnitude constraint in Fourier space can be defined in terms of the projection operator \cite{bauschke2002phase} as:
\begin{align}\label{PM_FD}
    \hat{P}_M \; \hat{\rho}(\boldsymbol{u}) & = \sqrt{I(\boldsymbol{u})} \frac{\hat{\rho}(\boldsymbol{u})}{|\hat{\rho}(\boldsymbol{u})|}, \hspace{1cm} \boldsymbol{u} \in Z \\ \nonumber
    & = \hat{\rho}(\boldsymbol{u}), \hspace{2.4cm} \boldsymbol{u} \notin Z.
\end{align}
Here, $\hat{P}_M$ is the Fourier modulus projection operator in Fourier domain and $Z$ is the set of pixels in $I(\boldsymbol{u})$ getting zero photon counts either due to noise or detector beam stop. The above equation can be translated in the real space as:
\begin{align}\label{PM_RD}
    {P}_M \; {\rho}(\boldsymbol{r}) = \mathcal{F}^{-1} \{ \hat{P}_M \hat{\rho}(\boldsymbol{u})\},
\end{align}
where, $P_M$ is the Fourier projection operator in object space. The second projection operator in object domain is the support projection $P_S$ which signifies the extent of the object contained within the support S and is defined as:
\begin{align}\label{PS}
    P_S \; \rho(\boldsymbol{r})& = \rho(\boldsymbol{r}), \hspace{1cm} \boldsymbol{r} \in S \\ \nonumber
                            &= 0, \hspace{1.5cm} \boldsymbol{r} \notin S
\end{align}
Classically the approach of phase retrieval methods is to reach the solution iteratively that satisfies the Fourier magnitude and \textit{a priori} object domain constraints altogether. Inspired by the work of Gerchberg and Saxton \cite{gerchberg1972practical} from electron microscopy for two measurement, the error reduction(ER) algorithm \cite{fienup1982phase} tries to solve the phase retrieval problem for single Fourier measurement via following solution update:
\begin{align}\label{ER}
    \rho_{n+1}(\boldsymbol{r}) = P_S P_M \; \rho_{n} (\boldsymbol{r}). 
\end{align}
The Fourier constraint being non-convex in nature makes the problem harder and the ER algorithm suffers from stagnation. The hybrid input output (HIO) algorithm \cite{fienup1978reconstruction} addresses the stagnation issue by introducing a negative feedback outside the object support. The HIO feedback step can be stated as:
\begin{align}\label{HIO}
    \rho_{n+1}(\boldsymbol{r}) & = P_M \; \rho_{n} (\boldsymbol{r}), \hspace{2.5cm} \boldsymbol{r} \in S \\ \nonumber
                           & = \rho_n (\boldsymbol{r}) - \beta P_M \rho_{n} (\boldsymbol{r}), \hspace{1cm} \boldsymbol{r} \notin S
\end{align}
where, $\beta$ is the feedback parameter in the range $(0,1)$. For noiseless data case, in a wide variety of images the HIO algorithm shows a significantly better convergence to an ideal solution which made it the workhorse for many imaging applications till date \cite{ekeberg2015three}. However, for the noisy data case, HIO solution also shows stagnation artifacts \cite{fienup1982phase} that corrupt the quality of the recovered solution. Phase retrieval with noisy data is an ill-posed problem, and therefore a unique solution to the problem cannot be found reasonably by satisfying both the Fourier magnitude (${P}_M$) and object support (${P}_S$) constraints. The difference map (DM) proposed by Elser \cite{elser2003phase} is an interesting approach which avoids local minima by 
using relaxed projections. The $(n+1)$-th iteration of the DM is given as:
\begin{align}\label{DM}
\rho_{n+1}(\boldsymbol{r}) = \rho_{n}(\boldsymbol{r}) + \beta(P_S R_{M}^{\gamma_M} - P_M R_{S}^{\gamma_S}) \rho_{n}(\boldsymbol{r}).
\end{align}
Here $R_{M}^{\gamma_M} = [(\hat{1}+\gamma_M)P_M - \gamma_M]$ and $R_{S}^{\gamma_S}= [(\hat{1}+\gamma_S)P_S - \gamma_S]$ are the relaxed generalized projections of $P_M$ and $P_S$ respectively \cite{stark1998vector} and $\hat{1}$ is the identity operator. $\beta$ is the relaxation parameter which needs to be optimized depending on the problem at hand while the other two parameters $\gamma_M$ and $\gamma_S$ are generally chosen to be $1/\beta$ and $-1/\beta$ for the optimal convergence \cite{elser2003random}. Another approach that is also known to surpass the stagnation problems is the relaxed averaged alternating reflections (RAAR)\cite{luke2004relaxed} algorithm, which is a descendant of the hybrid projection reflection (HPR) algorithm \cite{bauschke2003hybrid}. The RAAR algorithm shows rapid convergence to an optimal solution even when an exact solution satisfying both the Fourier magnitude and the support constraints does not exist. The RAAR update is given as:
\begin{align}\label{RAAR}
 \rho_{n+1}(\boldsymbol{r}) = [\frac{\beta}{2}(R_S R_M + \hat{1}) + (1-\beta)P_M] \rho_n(\boldsymbol{r}),
\end{align}
where, $R_S$ and $R_M$ are the reflections of $P_S$ and $P_M$ that are obtained by setting $\gamma_S=\gamma_M=1$. Apart from this, some other variants of HIO algorithm have also been reported for the noisy data case \cite{martin2012noise, rodriguez2013oversampling, trahan2013mitigating}. At this point we remark that among all the phase retrieval algorithms HIO, DM, and RAAR are the most widely used ones in CXI and have been tested for a numerous coherent X-ray imaging data sets. From application point of view where there is a wide range of CXI data, it is difficult to compare these algorithms 
. One of the reasons is that the behaviour of phase retrieval solution is not clearly understood for the various phasing schemes. Secondly, in order to evaluate the performance of these algorithms a metric can not be easily defined. 

In this paper our effort is to understand the nature of phase retrieval solution from a new perspective of the complexity parameter. The complexity information which can be computed \textit{a priori} and provides a measure of fluctuations present in the ground-truth object, is utilized to evaluate the characteristics of the phase retrieval solution. For the present work, we examine the behaviour of the solution obtained by the RAAR algorithm which is popularly used by the CXI community. Often in many CXI applications, it is a common practice to conclude the standard algorithms like HIO or RAAR with ER iterations to further stabilize/refine the RAAR solution and improve the error performance in Fourier domain \cite{chapman2006high, van2015imaging}. Therefore we use the combination of RAAR algorithm followed by 100 iterations of ER throughout the paper which will be referred by the name `RAAR-ER' algorithm.

For the sake of completeness, we start by describing the complexity parameter $\zeta_0$ which was introduced in \cite{butola2019phase}. As stated above, $\zeta_0$ is a measure of fluctuations present in pixel values of an object $\rho(x, y)$, where the co-ordinate \textbf{r} is expanded as $(x,y)$ . Mathematically the complexity is defined as: 
\begin{equation}\label{Complexity}
  \zeta_0 = \sum_{(x,y)} ( |\nabla_{x} \rho (x,y)|^2 + |\nabla_{y} \rho (x,y)|^2 ) .  
\end{equation} 
Here, $\nabla_x$ and $\nabla_y$ are the partial derivatives with respect to the $x$ and $y$ co-ordinates respectively. Using Parseval's theorem and employing the  derivative property of Fourier transform along with the notion of modified wave-numbers, the above equation can be equivalently written as:
\begin{equation}\label{complexity_mwn}
     \zeta_0 = \sum_{m=all pixels} \big[ \frac{\sin^2(2\pi f_{xm}\Delta x)}{(\Delta x)^2} +  \frac{\sin^2(2\pi f_{ym}\Delta y)}{(\Delta y)^2} \big] \; |\hat{\rho}(f_{xm}, f_{ym})|^2,
 \end{equation}
where $\Delta x$, $\Delta y$ are the grid spacing in spatial domain and $(f_{xm}, f_{ym})$ denote the spatial frequency samples as per the FFT convention. Note that the usual continuous case of this result is evident by setting $\Delta x, \Delta y \rightarrow 0$. 
Interestingly, the complexity parameter which is a real space constraint but derived directly from the raw Fourier intensity data $I(f_x, f_y) = |\hat{\rho}(f_{x}, f_{y})|^2$. 
Moreover, the complexity information in the form of $\zeta_0$ as defined in Eq. (\ref{complexity_mwn}) is readily available \textit{a priori} to all the phase retrieval algorithms. We observe empirically that the complexity parameter is not too sensitive to the typical Poisson noise in the Fourier magnitude data. This is because, relatively, the main contribution to $\zeta_0$ seems to be coming from the pixels with high Fourier intensity. Such pixels are generally minimally affected by Poisson noise. As discussed in Section 3, we will explain how the complexity information may be utilized when applying object domain constraints to obtain a noise robust phase retrieval solution.

\begin{figure}
\centering
\includegraphics[width=0.7\textwidth]{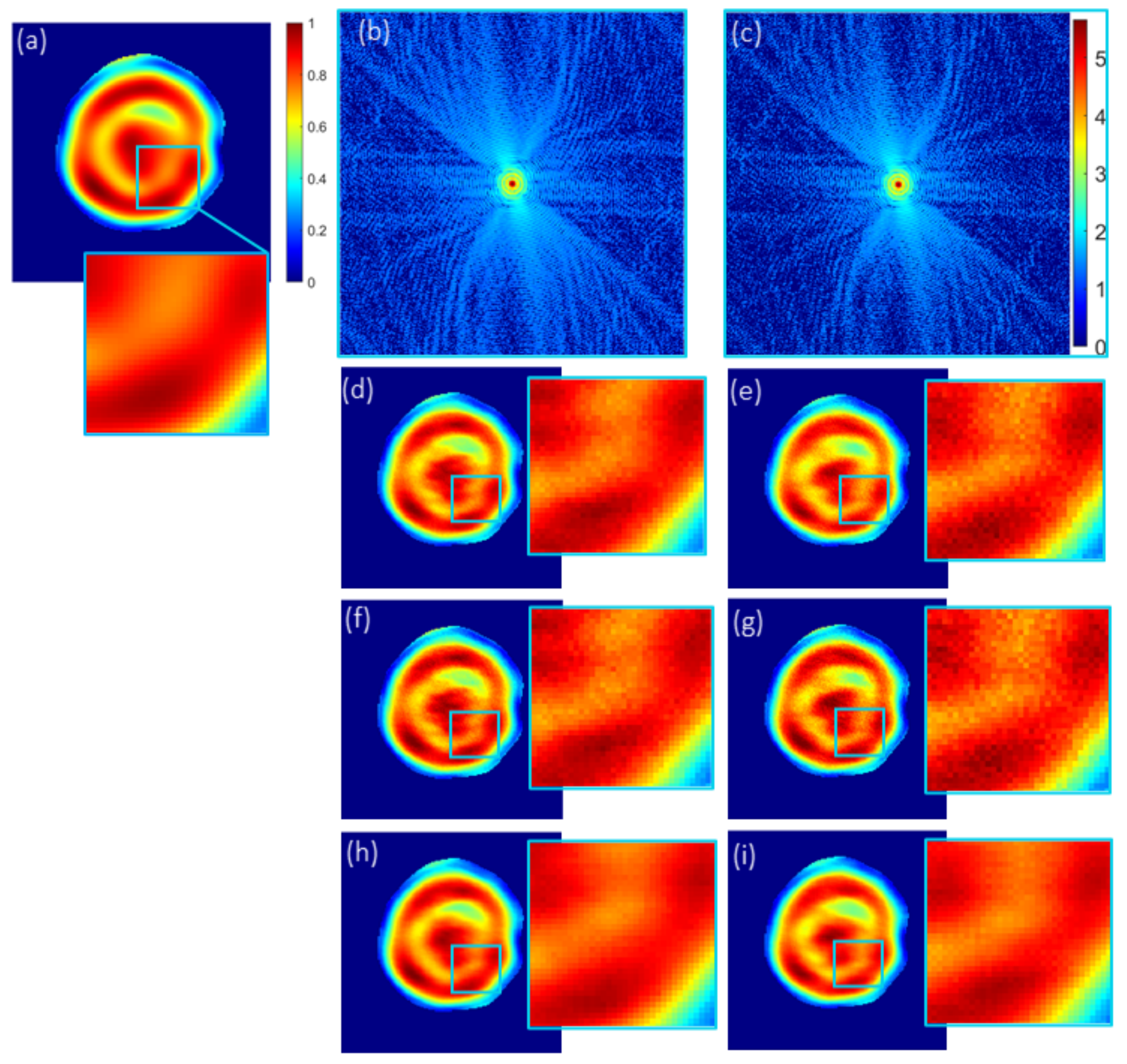}
\caption{Reconstruction from noisy diffraction data using RAAR-ER and CG-RAAR algorithms. \textbf{(a)} Ground-truth red blood cell(RBC) object. Simulated noisy oversampled Fourier intensity data of RBC in (a) corresponding to average light levels of \textbf{(b)} 1000 photons/pixel and \textbf{(c)} 500 photons/pixel. The data are shown as $|G|^{0.1}$ for the display purpose. Solutions recovered after \textbf{(d), (e)} 1000 iterations of RAAR alone \textbf{(f), (g)} concluding the solution in (d) and (e) by 100 ER steps and \textbf{(h), (i)} 1000 CG-RAAR iterations from the noisy data shown in (b), (c) respectively. For both the noisy data cases the CG-RAAR solution can be clearly seen to be free from artifacts. All the solutions are displayed in colorbar range [0,1].}
\end{figure}

\section{Complexity behaviour of the RAAR-ER solution with noisy data}
In this section we investigate the nature of a single run of RAAR-ER solution recovered from simulated noisy Fourier intensity data. The CXI data is noisy and has missing information due detector beam stop. However, for simplicity we restrict our initial simulations to noisy data without any beam stops. The missing pixels in the data due to beam stop requires additional discussion which will be described in the Section 4 when dealing with the experimental CXI data. 

In order to illustrate the effect of noise on the phase retrieval solution, we simulate two Fourier intensity data corresponding to a red blood cell (RBC) object in a computational window of size $512 \times 512$ pixels. The amplitude of this test object (central $160\times 160$ pixels) is shown in Fig. 1(a). The complex-valued RBC object for simulating noisy diffraction patterns was generated using a digital holographic microscope system (make: Holmarc) and corresponds to image plane field for the cell recovered using an an off-axis hologram recorded with a 650 nm laser. The noisy Fourier intensity data corresponding to this RBC object have been simulated by assuming the Poisson noise corresponding to two different average photon count levels of $1000$ and $500$ photons/pixels. The noisy data frames have $\sim 36\%$ and $51\%$ pixels with zero photon counts as shown in Figs. 1(b) and (c) respectively. These noise levels are comparable to those in the experimental CXI data to be discussed in Section 4. The Fourier intensities are over-sampled by a factor greater than $2$ since the support of the RBC object is small compared to the computational window.

We ran $1100$ iterations of RAAR-ER algorithm comprising of 1000 RAAR iterations, as per Eq. (\ref{RAAR}), that were concluded by $100$ ER iterations starting with the data in Fig. 1(b), (c). The relaxation parameter for RAAR update is selected to be $\beta= 0.9$. Since the object is complex-valued, only support constraint has been used. Figures 1 (d), (e) are the absolute magnitudes of the solution after 1000 RAAR iterations for the noisy data in Figs. 1(b) and (c) respectively, whereas the same solutions after additional 100 ER iterations are shown in Figs. 1(f), (g) respectively. From the solution recoveries we clearly see that for the higher noise data, both the RAAR and RAAR-ER solutions contain higher degree of grainy artifacts. 

Next we examine the complexity behaviour of RAAR and RAAR-ER solutions shown in Figs. 1(d)-(g). To do that in each RAAR or ER iteration, the complexity of the solution is calculated using Eq. (\ref{Complexity}) and is denoted by $\zeta$. In a similar vein, complexity inside $\zeta_{in}$ and outside $\zeta_{out}$ the support can be evaluated by summing over corresponding pixels. Clearly we have $\zeta = \zeta_{in} + \zeta_{out}$. Figure 2 (a) and (b) show the graphs of complexities $\zeta$, $\zeta_{in}$, $\zeta_{out}$ along with the ground-truth complexity $\zeta_0$ against the iteration number for the Fourier intensities with Poisson noise corresponding to average light levels of $1000$ photons/pixel and $500$ photons/pixel respectively. The complexities $\zeta$ and $\zeta_{out}$ for both the cases stabilize to values higher and lower than the ground-truth complexity $\zeta_0$. The outside support complexity $\zeta_{out}$ is seen to stabilize to a slightly higher value for the higher noise data. Interestingly, in terms of complexity behaviour of solution, we observe that for the noisier data case the solution's complexity within the support $\zeta_{in}$ stabilizes at a higher level for data with higher noise. The manifestation of higher complexity or fluctuations inside the support region with increasing noise can be clearly seen in solutions in Figs. 1(e), (g) when compared to those in Figs. 1(d), (f). When ER iterations are applied after 1000 RAAR iterations, $\zeta_{out}$ stabilizes to a lower value than that in the RAAR case. On the other hand $\zeta_{in}$ increases as compared to that at the end of RAAR iterations and also contributes most to the total complexity $\zeta$. The reduction of $\zeta_{out}$ during ER iterations is easy to understand as the pixels outside the support region are set to zero during the ER iterations, so that, only contribution to $\zeta_{out}$ comes from the pixels adjacent to the support boundary. From the complexity curves $\zeta_{in}$, $\zeta_{out}$ in Figs. 2(a), (b) and solutions in Figs. 1(f), (g) we see that in the process of improving the Fourier domain error by applying ER steps, the complexity of the RAAR solution within the support region has increased. This increased complexity causes additional grainy artifacts that in our opinion are due to over-fitting of noise in the data. The application of ER iterations that typically reduce the data domain error is thus not necessarily desirable in our opinion, as this is accompanied by appearance of increased grainy artifacts within the support. The grainy artifacts are clearly spurious when the solution is compared with the ground-truth object in Fig. 1 (a). The high frequency nature of the grainy artifacts may also lead to a slightly exaggerated estimate of the resolution of the solution based on the PRTF curve as we will see later in Section 5.

The prior literature \cite{martin2012noise,rodriguez2013oversampling,guo2015iterative} has stated the significance of outside support region in relation to noise in the Fourier intensity data. However, the procedures for suppressing outside support information in the literature are somewhat \textit{ad hoc} in our opinion. In the complexity guided phase retrieval approach described next, the \textit{a priori} knowledge about the ground-truth complexity $\zeta_0$ plays an important role in application of constraints in the object domain (both within and outside the support).

\begin{figure}
\centering
\includegraphics[width=0.7\textwidth]{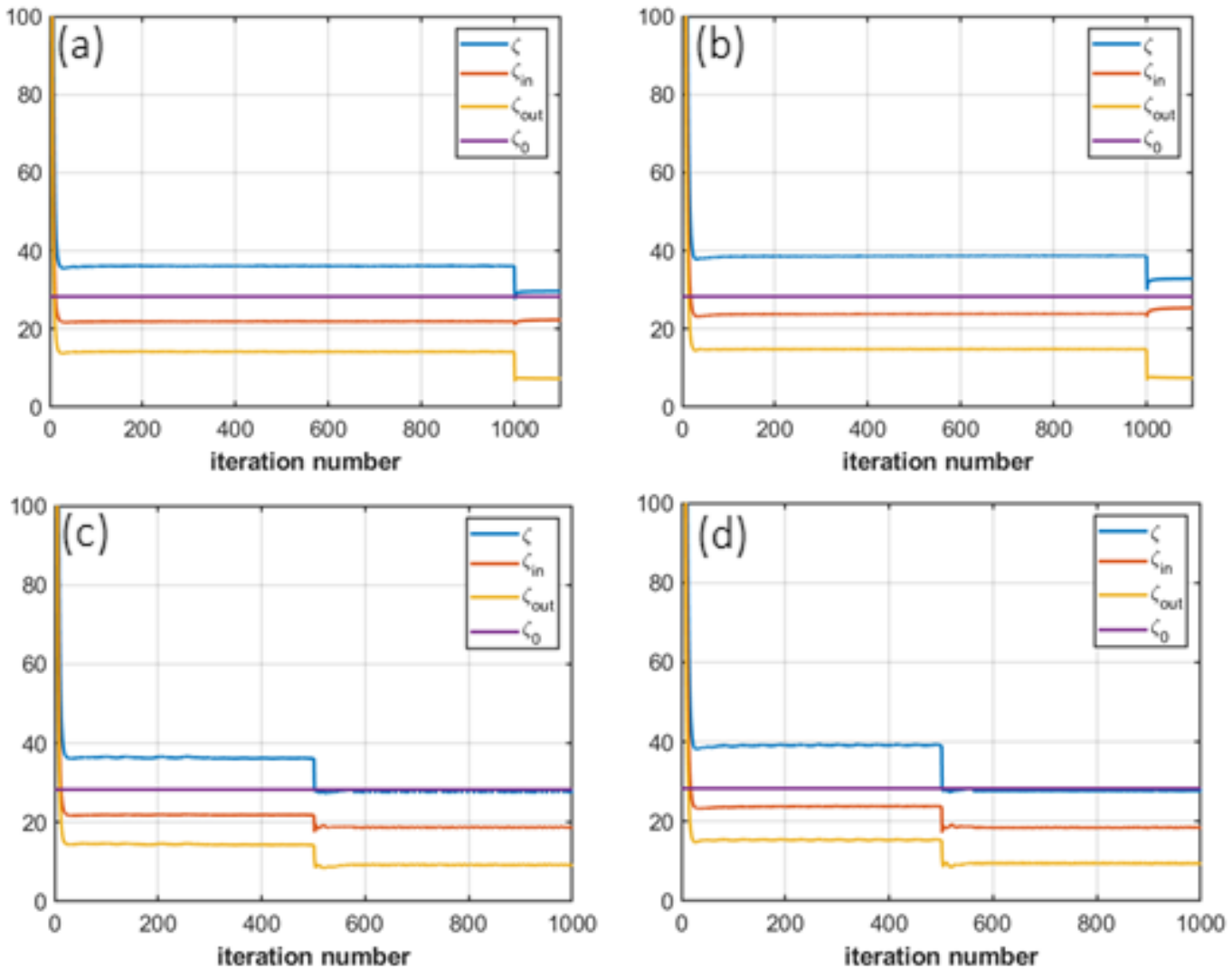}
\caption{ Behaviour of complexity of solution with iteration number for \textbf{(a), (b)} RAAR-ER and \textbf{(c), (d)} CG-RAAR algorithm for the average light level of 1000 photons/pixel and 500 photons/pixel respectively.}
\end{figure}

\section{Complexity-guided RAAR (CG-RAAR) algorithm}
From the complexity curves in Figs. 2 (a), (b), we observed that the solution complexity $\zeta$ for RAAR as well as RAAR-ER algorithms stabilizes to a higher numerical value compared to the ground-truth complexity $\zeta_0$ computed from the raw data. The concept of complexity-guidance is based on the strategy of matching $\zeta$ with $\zeta_0$. Instead of concluding the RAAR algorithm with ER iterations (which leads to undesirable increase in solution complexity within the support), we match the total complexity $\zeta$ of the guess solution with the ground-truth complexity $\zeta_0$ in each RAAR iteration. To bring down $\zeta$ to the level of $\zeta_0$, the values of $\zeta_{out}$ and $\zeta_{in}$ are decreased via complexity reduction steps in an adaptive manner. In our previous work \cite{butola2021complexity}, while studying the behaviour of HIO solution, we observed that for the increasing noise in the data, the $\zeta_{out}$ showed a monotonic increase with iteration number and for that reason $\zeta_{out}$ was required to be suppressed actively in each HIO iteration. Unlike HIO solution, the RAAR solution does not show continuously increasing complexity outside the support region with iterations and the contribution of $\zeta_{in}$ to the total complexity $\zeta$ is more as compared to $\zeta_{out}$. Therefore, most complexity reduction in the present case needs to be done within the support region. It is interesting to note that an improvement over the native RAAR algorithm has been suggested previously by introducing a generalized feedback term inside the support region \cite{adams2012generalization}. Based on our observations we control the total complexity of the RAAR solution primarily by suppressing the complexity within the support which is essentially like a complexity feedback within the support. 

We remark that complexity-guidance is essentially a regularization scheme where complexity (or degree of fluctuations) in the desired solution is pre-determined from the raw data. In order to implement complexity-guidance the recursive iterations for complexity reduction are applied to the RAAR solution $\rho_{n+1}(\boldsymbol{r})$ in Eq. (\ref{RAAR}) in the outside and inside the support region  in each outer RAAR iteration. For brevity of notation we denote the outside and inside part of the solution as $\rho_{out}$ and $\rho_{in}$ respectively. Firstly, recursive sub-iterations are applied outside the support as represented by: 
\begin{align}\label{eq:tv2_red_1}
        \rho_{out}(\boldsymbol{r}) \longleftarrow \rho_{out}(\boldsymbol{r}) - \tau ||\rho_{out}(\boldsymbol{r})||_{2} [\hat{u}]_{\rho_{out}}.
\end{align}
Next, further recursive complexity reduction iterations similar to above are also applied to the $\rho_{in}$: 
\begin{align}\label{eq:tv2_red_2}
        \rho_{in}(\boldsymbol{r}) \longleftarrow \rho_{in}(\boldsymbol{r}) - \tau ||\rho_{in}(\boldsymbol{r})||_{2} [\hat{u}]_{\rho_{in}}.
\end{align}
For both the above equations $\tau$ is the step size which was chosen to be $5 \times 10^{-3}$ so that the solution changes by a small amount in each sub-iteration and $\hat{u}$ is the normalized functional gradient of complexity defined as: 
\begin{align}\label{unit vector}
         \hat{u} &= \frac{\nabla_{\rho^*} \zeta(\rho, \rho^*)}{|| \nabla_{\rho^*} \zeta(\rho, \rho^*) ||_2},
 \end{align}
 where, $ \nabla_{\rho^*} \zeta(\rho, \rho^*) = - \nabla^2 \rho$ is the complex derivative \cite{brandwood1983complex} of complexity parameter $\zeta$. As explained in \cite{butola2021complexity}, the number of sub-iterations for $\zeta_{out}$ reduction are adjusted by actively tracking the trend of $\zeta_{out}$. Particularly, the value $\zeta_{out}$ is nudged towards the mean minus the standard deviation of $\zeta_{out}$ of the previous $20$ iterates in order to control any growth in $\zeta_{out}$. After that, the complexity reduction sub-iterations within the support are applied till the numerical value of the total complexity $\zeta_{in} + \zeta_{out}$ is close to (within $0.5 \%$) the ground-truth complexity $\zeta_{0}$. The number of sub-iterations in this step typically depends on the noise level (higher noise data requires more complexity reduction sub-iterations inside the support). The (n+1)-th iterate of CG-RAAR is updated as:
\begin{align}
    \rho_{n+1}(\boldsymbol{r}) &= \rho_{in}(\boldsymbol{r}), \hspace{1cm} \boldsymbol{r} \in S \notag \\
         &= \rho_{out}(\boldsymbol{r}),\hspace{1cm} \boldsymbol{r} \notin S.
\end{align}
Here $\rho_{in}$ and $\rho_{out}$ refer to the inside and outside support solutions after applying the complexity guidance sub-iterations. The recovered solutions using CG-RAAR from the same noisy data in Figs. 1(b),(c) are shown in Figs. 1(h), (i) respectively. Visually, for the higher noisy data case the CG-RAAR solution in Fig. 1(h), (i) are free from severe grainy artifacts that were seen in RAAR and RAAR-ER solutions in Figs. 1(e), (g) respectively. 

Our implementation used initial $500$ RAAR iterations followed by $500$ RAAR iterations with complexity guidance. In Figs. 2(c), (d), while observing the complexity behaviour of the CG-RAAR solution with iterations, we see that after $500$ iterations when complexity-guidance sub-iterations are applied, by design the complexity of the solution matches with $\zeta_0$. In particular the high frequency artifacts within the support which arise due to the noise in the Fourier intensity data are suppressed by the complexity-reduction steps. Therefore, the CG-RAAR solution is smoother in nature and is much closer to the ground-truth shown in Fig. 1(a). We want to emphasize here that all the results shown in Fig. 1 correspond to a single run of the corresponding algorithms (RAAR or RAAR-ER or CG-RAAR). The computational time taken by the $1100$ RAAR-ER iterations is $\sim 13$ seconds while $1000$ iterations of CG-RAAR algorithm takes $\sim 3$ minutes due to complexity reduction sub-iterations. Despite more computational effort, the phase retrieval with complexity-guidance provides much reliable and robust phase retrieval solution. As it is known that phase retrieval algorithms are formulated to find a better estimate if more constraints are satisfied. Therefore, for the practical scenario with noisy data this new constraint complexity provides a better estimate for phase retrieval. Additionally, the trial solutions selected for the averaging process can be made better by methods suggested by \cite{fienup1986phase} such as voting, patching, use of non-centrosymmetric support etc.\cite{chapman2006high}. Here, the issues of stripes like artifacts or twin-stagnation (arising due to phase inaccuracies) are easily addressed by sparsity enhancing functionals like complexity, total variation, Huber or other penalties \cite{gaur2015sparsity}.

\begin{figure}
\centering
\includegraphics[width=0.7\textwidth]{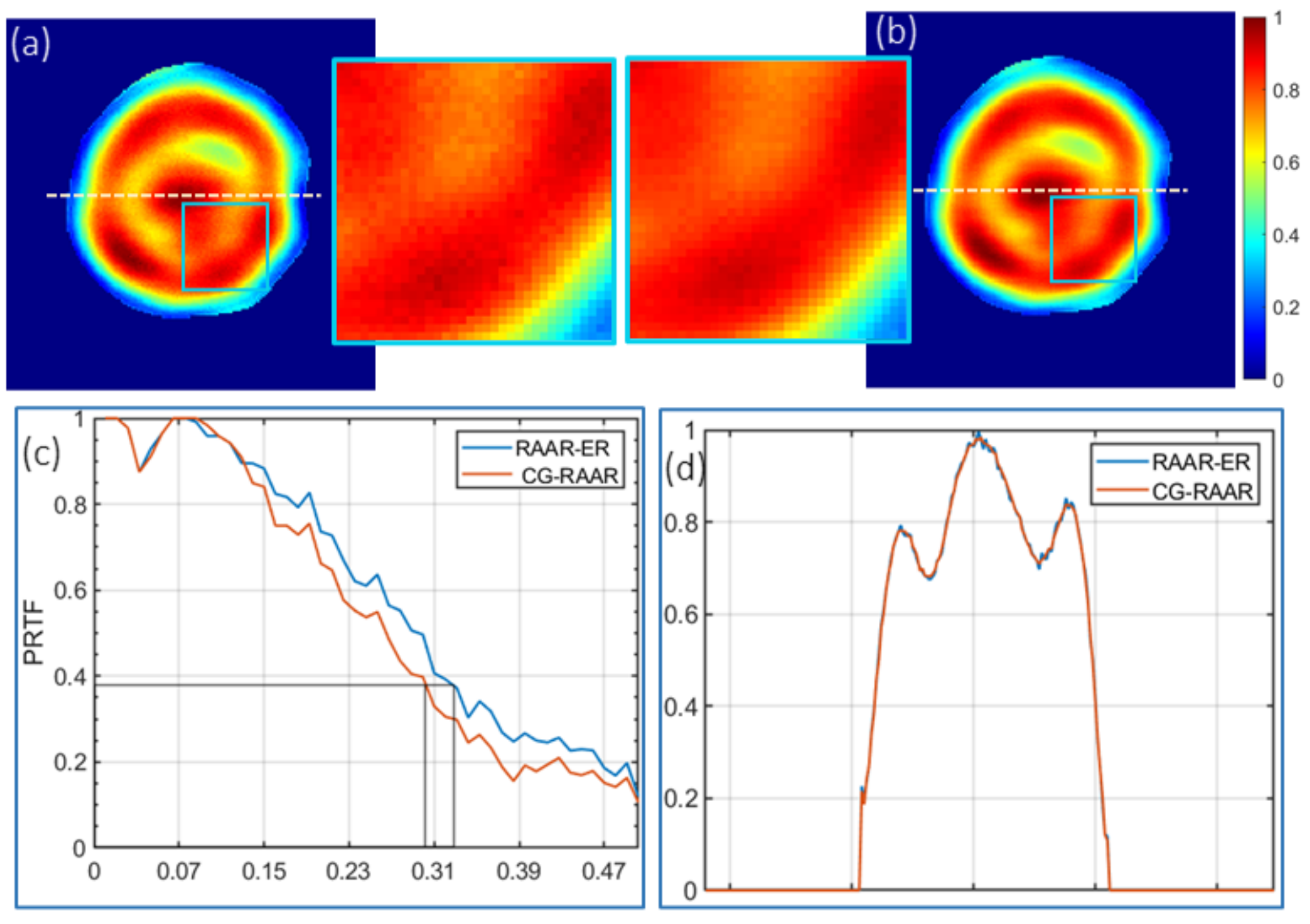}
\caption{ The solution recovery after the averaging over 10 trial solutions of \textbf{(a)} RAAR-ER algorithm and \textbf{(b)} CG-RAAR algorithm for the noisier data case of 500 photons/pixel. The solutions are displayed in colorbar [0,1]. \textbf{(c)} PRTF vs spatial frequency curve and \textbf{(d)} is profile across the yellow dotted lines in the solutions displayed in (a), (b). For both the solutions the recovered spatial resolution is $\sim$ 3 pixels. Clearly, from the profile plots in RAAR-ER the noisy fluctuations smaller than 3 pixels are clearly visible whereas the CG-RAAR profile is almost flat within 3 pixels}
\end{figure}
 
\section{Evaluating the resolution of reconstructed solution using phase retrieval transfer function (PRTF)}
The reconstruction of solution from the experimental Fourier transform intensity data consisting of noise and missing pixels due to detector blocking poses challenges. A single run of the algorithm starting with a random guess cannot be fully relied upon. It is therefore a standard practice to arrive at a better estimate of the solution by averaging a large number of trial solutions ~\cite{shapiro2005biological}. These independent trial solutions are obtained by running the phase retrieval algorithm over hundreds of runs that are initialized with a different random guess. Among a number of trial reconstructions only some following a certain criterion are selected for the averaging process. The averaging criterion can be decided on the basis of least errors in real space or Fourier space. Here we have selected correlation coefficient between trial solution and a reference solution with least error as an averaging criterion. Since the final reconstruction is based on the  averaging process, the goodness of reconstruction can be assessed using phase retrieval transfer function (PRTF) which compares the magnitude of Fourier transform of averaged complex reconstruction with the measured Fourier modulus and defined as \cite{chapman2006high, latychevskaia2018iterative}:
\begin{align} \label{PRTF}
    PRTF(\boldsymbol{u}) = \frac{| \hat{\rho}_{avg}(\boldsymbol{u})|} {\sqrt{I(\boldsymbol{u})}}.
\end{align}
Here, the average of complex Fourier estimates $ \hat{\rho}_{avg} (\boldsymbol{u})$ can be simplified as :
\begin{align}
\hat{\rho}_{avg} (\boldsymbol{u}) &= \frac{1}{T} \sum_{t = 1}^T \hat{\rho}_t (\boldsymbol{u}) \\ \notag
                                & = \frac{1}{T} \sum_{t = 1}^T \mathcal{F} \{ \rho_t(\boldsymbol{r})\} \\ \notag
                                & = \frac{1}{T}  \mathcal{F} \{ \sum_{t = 1}^T  \rho_t(\boldsymbol{r})\} ,
\end{align}
where $t = 1, 2, ....., T$ denotes the trial number index and $\rho_t(\boldsymbol{r})$ represents the phase adjusted trial reconstructions following the correlation criterion for averaging. In order to add the trial solutions coherently the phases of trial solutions are adjusted to a common value $\phi_0$ so that the variation in phases does not reduce the average value. The phase adjustment of all the trial reconstructions can be done as per \cite{chapman2006high}. The PRTF measure can also provide the resolution of the recovered solution by selecting the point on the plot of radial averaging of PRTF vs. radial spatial frequency where PRTF falls to 1/e \cite{chapman2006high}. In this section, we first present the PRTF curves for the simulated data followed by similar analysis of the experimental data in the next section.

Figures 3(a) and (b) are the solutions obtained after averaging over $T = 10$ trials of RAAR-ER and CG-RAAR algorithms respectively for noisy data in Fig. 1(b). The correlation coefficient with value 0.96 is chosen as the averaging criterion. For the simulated data, we have not assumed any missing pixels due to detector blocking and assumed only Poisson noise corrupted data. In this case we observe that an average over a small number of trial solutions is sufficient to stabilize the PRTF curve.
One can clearly see that the averaging process has certainly reduced the grainy appearance of the RAAR-ER solution. However, some of the faint artifacts are still persistent which may possibly get misinterpreted as the real features, particularly in the case of experimental Fourier intensity data where the ground-truth object is unknown. Since the individual CG-RAAR solution was already free from grainy artifacts, the averaged CG-RAAR solution closely resembles the ground-truth object. The performance of the CG-RAAR and RAAR-ER solutions are tabluated in Table 1 in terms of error in Fourier and real domain. The real space error metric $E_R$ is defined as \cite{fienup1997invariant}:
\begin{align}
    E_R^2 = min( E^2(\rho_{est}(\boldsymbol{r})), E^2(\rho_{est}^*(-\boldsymbol{r})) ), 
\end{align}
where,
\begin{align}
    E^2(\rho_{est}(\boldsymbol{r})) = \frac{\sum |\rho_{est}|^2 + \sum |\rho|^2 -2corr(\rho_{est}, \rho)}{\sum |\rho|^2}.
\end{align}
Here, $\rho_{est}$ represents the solution estimate (individual trial or average solution), $\rho$ is the ground-truth object, and $corr(\rho_{est}, \rho)$ is the correlation between $\rho_{est}$ and $\rho$. Next the relative Fourier domain error $E_F$ for the Fourier modulus estimate $\hat{\rho}_{est}(\boldsymbol{u})$ with respect to the measured Fourier modulus $\sqrt{I(\boldsymbol{u})}$ can be written as:
\begin{align}
    E_F^2 = \frac{|| \sqrt{I(\boldsymbol{u})} - |\hat{\rho}_{est}(\boldsymbol{u})| \;  ||_2}{|| \sqrt{I(\boldsymbol{u})}||_2}.
\end{align}
Both RAAR-ER and CG-RAAR solutions seem to have similar performance in terms of the above metrics. 

Figure 3(c) shows the PRTF curves for the solutions in Figs. 3(a), (b) obtained by both the algorithms. The spatial resolution for the averaged RAAR-ER and CG-RAAR solutions is $\sim 3$ pixels. The resolution estimate for the RAAR-ER solution is slightly better compared to the CG-RAAR solution. This higher resolution estimate is however spurious and arises mainly due to the remaining grainy artifacts. From the zoomed-in portions of Figs. 3(a), (b) and the profile plot in Fig. 3(d) across the yellow dotted lines in both the solutions we see that the CG-RAAR solution is smooth on a $3$ pixels scale unlike the RAAR-ER solution. Therefore, it is quiet evident from the PRTF plot that RAAR accompanied by complexity-guidance leads to a self-consistent solution than that obtained by the traditional methodology of RAAR concluded with ER algorithm for noisy Fourier intensity data. 

\begin{table}
 \caption{Performance of single trial and averaged RAAR-ER and CG-RAAR solutions for the Fourier intensity data with average photon counts of $500$ photons/pixel in terms of real space error ($E_R$) and Fourier domain relative error ($E_F$)}
  \centering
  \begin{tabular}{lll}
    \toprule
     \textbf{Solution} & $\boldsymbol{E_R}$    & $\boldsymbol{E_F}$             \\
    \midrule
    RAAR-ER (Individual)   & $3.03e-3$      & $1.83e-4$           \\
 CG-RAAR (Individual)   & $2.86e-3$     & $3.29e-4$         \\
 RAAR-ER(Average)       & $2.57e-3$      & $1.23e-3$        \\
 CG-RRAR(Average)      & $2.66e-3$     & $1.38e-3$        \\
    \bottomrule
  \end{tabular}
  \label{tab:table}
\end{table}

\begin{figure}
\centering
\includegraphics[width=0.95\textwidth]{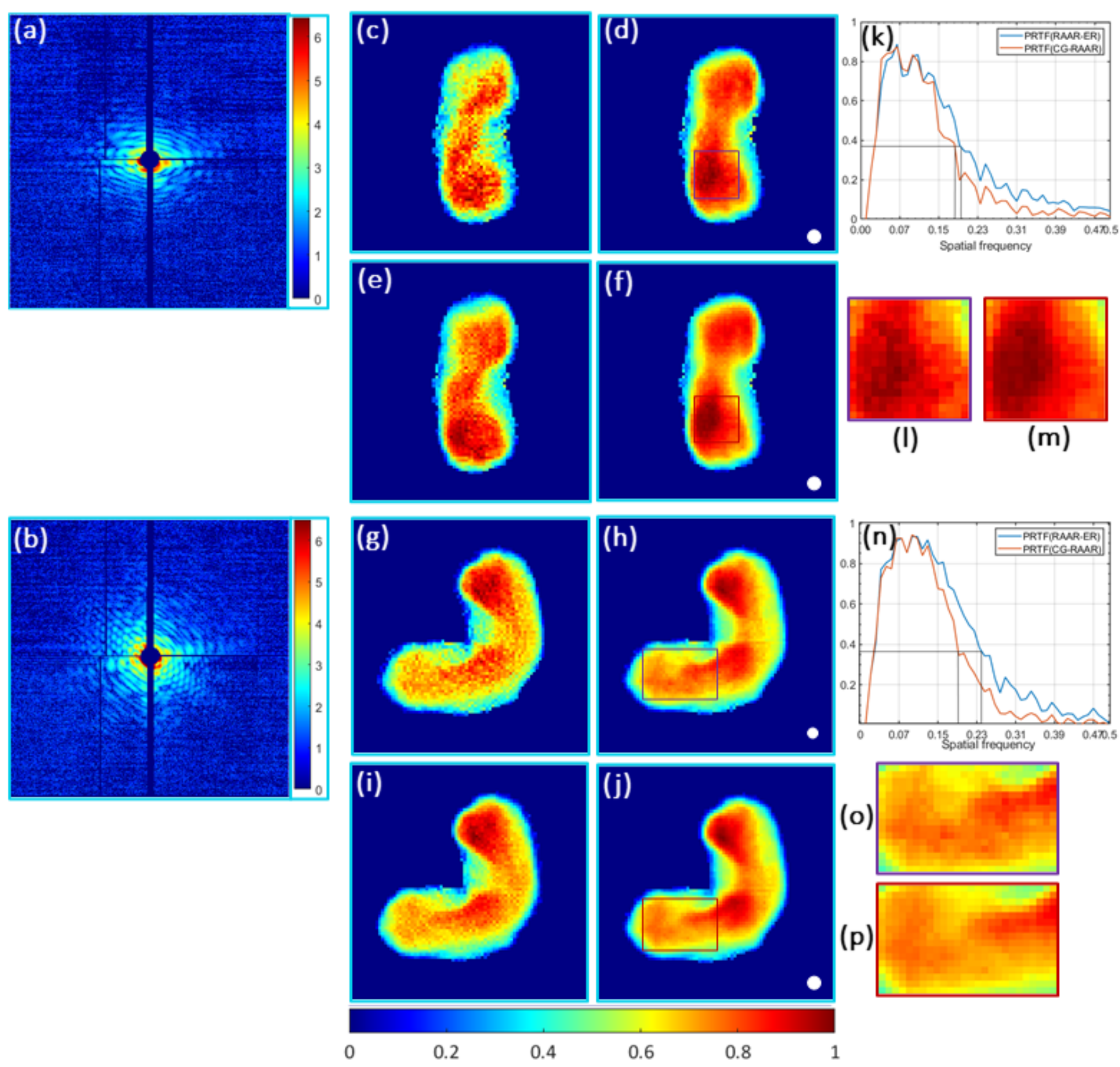}
\caption{Performance of RAAR-ER and CG-RAAR when applied to the experimental data. \textbf{(a), (b)} Noisy and incomplete coherent X-ray imaging diffraction data. \textbf{(c), (e)} individual and \textbf{(d), (f)} average solutions using RAAR-ER and CG-RAAR algorithms respectively for the data in (a). \textbf{(g), (h)} Individual and \textbf{(i), (j)} average solutions using RAAR-ER and CG-RAAR algorithms respectively for the data in (b). All the solutions are shown in the scale of $[ 0, 1 ]$. \textbf{(k), (n)} are the respective PRTF versus radial spatial frequency plots for the average solutions. The purple and red boxes in the RAAR-ER and CG-RAAR solutions are shown in \textbf{(l), (o)} and \textbf{(m), (p)} respectively as their zoom-in views. The white circles represents the resolution. }
\end{figure}

 \section{Experimental Results}
In order to further demonstrate the performance of CG-RAAR algorithm, we take an experimental CXI data submitted to the coherent X-ray imaging data bank (CXIDB) \cite{maia2012coherent}. The experiment was carried out in Linac coherent light source (LCLS) and the extensive details about the experimental set-up are described in ~\cite{van2015imaging}. The data set contains ten Fourier intensity frames corresponding to live cyanobacteria at ten different stages. However, for the illustration purpose we take only two intensity data (Figs. 2(i), (j) of ~\cite{van2015imaging}), which contain a total of $1.6 \times 10^6$ and $8.5 \times 10^5$ photon counts and are shown in Figs. 4(a), (b) respectively. The Fourier intensities in Figs. 4(a) and (b) have $\sim 51 \%$ and $46\%$  missing pixels due to detector blocking and Poisson noise. The masks provided in the `.cxi' file for the corresponding intensities were used as the known object supports. The experimental data has blocked pixels and the only modification required to handle the blocked pixels is to replace them with the Fourier magnitude corresponding to the current iterate. We ran both RAAR-ER and CG-RAAR algorithms 400 and 200 times independently in a MATLAB environment. In the first case, 2000 RAAR iterations were concluded by 100 ER steps. In CG-RAAR initial 1500 RAAR iterations were followed by the 500 outer RAAR iterations with complexity-guidance sub iterations. Since we are filling the blocked pixels with the Fourier magnitude of the current iterate, the ground-truth complexity estimate also needs to be revised in each CG-RAAR iteration. We however note that by $1500$ RAAR iterations (after which we use the complexity guidance), the numerical values in the blocked pixels are seen to stabilize and the estimated $\zeta_0$ does not fluctuate significantly. All these cases with multiple independent initial random guesses were ran on the locally available high power computing (HPC) cluster. Figures 4(c) and (d) show one of the trial solutions obtained using the RAAR-ER and CG-RAAR algorithms respectively. Visually, we observe that the single instance of the CG-RAAR solution is smoother compared to that obtained using the RAAR-ER algorithm. The averaged solution for both the respective algorithms are shown in Figs. 4(e), (f). The averaging procedure is the same as followed in the simulations. Based on the $96 \%$ correlation criterion as before, 261 out of 400 RAAR-ER solutions were selected for the averaging while in CG-RAAR 94 out of 200 solutions followed the criterion. We observe that the PRTF curve is stabilized for both RAAR-ER and CG-RAAR cases and additional trial runs will not have any perceptible effect on the PRTF curve. In order to assess the quality of averaged solutions, we plot the PRTF as a function of spatial frequency which is shown in Fig. 4(k). The PRTF curve is vanishing near zero frequency since there are blocked pixels in this region where the Fourier intensity data is missing. The spatial resolution (reciprocal of spatial frequency) at which the PRTF falls to $1/e$ is 4.54 and 4.76 for the RAAR-ER and CG-RAAR solution respectively which can be rounded off to $\approx 5$ pixels. From the zoomed-in portion of the averaged RAAR-ER solution in Fig. 4(l), we clearly see features smaller than 5 pixels that are spurious and difficult to interpret. On the other hand from Fig. 4(m) the CG-RAAR solution is clearly observed to be smoother and consistent with the spatial resolution reflected in PRTF curve. We remark that the spatial resolution given here is in terms of pixels however one can always express it in physical units by the relation $\Delta x = \frac{\lambda z}{\Delta x_{det} N}$, where $\lambda$: wavelength of the X-rays, z: distance between specimen and detector, $\Delta x_{det}$: pixel size of the detector, and N: half pixel number. For the Fourier intensity data shown in Fig. 4(b), one can make similar observations from Figs. 4(g)-(j) and (n)-(p). The final solutions shown in Figs. 4(h) and (j) are obtained after averaging over $385$ and $188$ RAAR-ER and CG-RAAR trial solutions respectively. Once again the averaged RAAR-ER solution still seems to retain some grainy artifacts that are not present in the CG-RAAR solution. The grainy artifacts in the RAAR-ER solution lead to a slightly better resolution estimate which is spurious in our opinion. CG-RAAR solution however seems to have features consistent with the resolution estimate provided by the PRTF curve. It is worth noting from our results that the CG-RAAR solution is artifact-free despite the fact that the number of trial solutions used for the averaging process in CG-RAAR case is much less than half the number of trial solutions used for RAAR-ER case. This is expected since even the single run of CG-RAAR solution has much reduced artifacts compared to that of the RAAR-ER solution. 
 
\section{Conclusion}
In conclusion, we have examined the nature of phase retrieval solutions obtained from the noisy Fourier diffraction data in CXI experiments. In particular, we studied the solution obtained by the well-known methodology of RAAR algorithm combined with ER iterations from a simulated noisy Fourier intensity data using complexity parameter. The complexity parameter is an object domain constraint which provides a measure of fluctuations present in the ground-truth object; and can be calculated \textit{a priori} from the raw Fourier intensity data. With increasing noise in the data the complexity of RAAR solution within the support stabilized to a higher level higher and the total solution complexity also followed a similar trend. When the concluding 100 ER steps were applied for the purpose of refining the RAAR solution, the resulting solution showed undesirable increase in the complexity of the solution inside the support region that was manifested in the form of severe grainy artifacts in the solution. In order to address this issue we proposed a complexity guided RAAR (CG-RAAR) algorithm which does not show appearance of such artifacts. The goodness of the algorithm and the recovered resolution are judged using the PRTF plot. The traditional RAAR-ER solution seems to retain some grainy artifacts (even after averaging over a number of trial solutions), whose feature size is smaller that the resolution estimated by the PRTF. The proposed CG-RAAR algorithm does not show these spurious features and has features consistent with the resolution estimate provided by the PRTF. The main idea behind complexity guidance is to match the complexity of RAAR solution with the desired complexity that is calculated from the raw Fourier intensity. It is important to note that the complexity-guidance is a regularization scheme which avoids over fitting of noise by using complexity information. The performance of CG-RAAR algorithm is demonstrated for the experimental CXI data submitted to the CXIDB which consists of missing pixels due to detector blocking along with the noise. 
It is worth highlighting that the number of individual CG-RAAR trial solutions required for the averaging process is less than the half number that are needed for RAAR-ER method. From our simulation and experimental results it is evident that the complexity-guidance idea incorporated to the traditional algorithms like RAAR(or HIO) can provide reliable and self-consistent solutions for practical CXI data with significantly reduced number of trials. Coupled with the new experimental developments in CDI, we believe that the idea of complexity-guidance can possibly benefit the field of X-ray coherent diffraction imaging.

\bibliographystyle{unsrt}  
\bibliography{references}

\end{document}